\documentclass[preprint,aps,nofootinbib]{revtex4}

\usepackage{graphicx}

\begin{document}

\title{Earth matter density uncertainty in atmospheric neutrino oscillations}
\author{Pei-Hong Gu}
\email{guph@mail.ihep.ac.cn} \affiliation{Institute of High Energy
Physics, Chinese Academy of Sciences, P.O. Box 918-4, Beijing
100049, People's Republic of China}
\begin{abstract}
That muon neutrinos $\nu_{\mu}$ oscillating into the mixture of
tau neutrinos $\nu_{\tau}$ and sterile neutrinos $\nu_{s}$ has
been studied to explain the atmospheric $\nu_{\mu}$ disappearance.
In this scenario, the effect of Earth matter is a key to determine
the fraction of $\nu_{s}$. Considering that the Earth matter
density has uncertainty and this uncertainty has significant
effects in some neutrino oscillation cases, such as the $CP$
violation in very long baseline neutrino oscillations and the
day-night asymmetry for solar neutrinos, we study the effects
caused by this uncertainty in the above atmospheric $\nu_{\mu}$
oscillation scenario. We find that this uncertainty seems to have
no significant effects and that the previous fitting results need
not to be modified fortunately.
\end{abstract}
\maketitle

To explain the atmospheric muon neutrinos $\nu_{\mu}$
disappearance, the scenario of $\nu_{\mu}$ oscillating into
$\nu_{+}$ has been studied\cite{sk,fogli,nakaya}, where $\nu_{+}$
is the mixture of tau neutrinos $\nu_{\tau}$ and sterile neutrinos
$\nu_{s}$, and defined as $\nu_{+} = \nu_{\tau} cos \xi + \nu_{s}
sin \xi$. Since for so-called "matter effects"\cite{msw}, the
oscillation probabilities $P_{\nu_{\mu} \rightarrow \nu_{\tau}}$
and $P_{\nu_{\mu} \rightarrow \nu_{s}}$ are different for a muon
neutrino with certain energy that travels a distance in Earth, one
can expect to give a limit on $\xi$. The reported results from
Super-kamiokande have given limits on $sin^{2} \xi
$\cite{sk,fogli,nakaya}.

In the calculation of $P_{\nu_{\mu} \rightarrow \nu_{s}}$, the
neutron number density of Earth is a critical quantity. However,
today the knowledge of Earth matter density which determines the
neutron number density is only to some certain
precision\cite{geller}. As to the preliminary reference Earth
model (PREM)\cite{prem}, the uncertainties due to the local
variation have been documented\cite{jeanlow}. Quantitatively its
precision is roughly $5 \%$ averaged per spherical shell with
thickness of 100 km or so\cite{bolt}.

The effects of Earth matter density uncertainty have been studied
in some neutrino oscillation cases, such as the $CP$ violation in
very long baseline neutrino oscillations\cite{shan1,shan3} and the
day-night asymmetry for solar neutrinos\cite{shan2}. Ones find
this uncertainty has significant effects in these
cases\cite{shan1,shan3,shan2}. Since the Earth matter is a key to
determine the fraction of $\nu_{s}$, this uncertainty could also
have an effect on the limit of $sin^{2} \xi$. In this brief
report, we study the density uncertainty in Earth matter and then
investigate its implications on the results of $sin^{2} \xi$.

We begin our discussions with the effective Hamiltonian that
governs the propagation of the neutrinos in matter. In the (2+2)
models\cite{barger}, the relevant $(\nu_{\mu},\nu_{+})$ evolution
is given by the Schr$ \ddot{o}$dinger equation

\begin{equation}
i \frac{d}{dx} \left(\begin{array}{cc}
\nu_{\mu} \\
\nu_{+}
\end{array}\right) \simeq H(x) \left(\begin{array}{cc}
\nu_{\mu} \\
\nu_{+}
\end{array}\right)
\end{equation}
with the effective Hamiltonian\cite{fogli}
\begin{equation}
\label{ham} H(x)= \frac{\Delta m^{2}}{4E_{\nu}}
\left(\begin{array}{cc}
\ - \cos2\theta & \sin2\theta \\
\ \sin2\theta & \cos2\theta
\end{array}\right)+ \sqrt{2} G_{F}\left(\begin{array}{cc}
\ 0 & 0 \\
\ 0 & \frac{1}{2} \sin ^{2} \xi N_{n}(x)
\end{array}\right)\ .
\end{equation}
Here $E_{\nu}$ is the neutrino energy, $\Delta m^{2}$ and $\theta$
are the usual mass and mixing parameters in the $\nu_{\mu}
\rightarrow \nu_{\tau}$ oscillating model, $G_{F}$ is the Fermi
constant, and $N_{n}(x)=\rho (x) N_{A} (1-Y_{e}(x))$ is the
neutron number density with $\rho (x)$ the matter density in
$\textrm{g}/\textrm{cm}^{3}$, $N_{A}$ the Avogadro number and
$Y_{e}(x)$ the electron number fraction, respectively. For
increasing values of $sin^{2} \xi$, we get a smooth interpolation
from $\nu_{\mu} \rightarrow \nu_{\tau}$ oscillations $(sin^{2} \xi
=0)$ to pure $\nu_{\mu} \rightarrow \nu_{s}$ oscillations
$(sin^{2} \xi =1)$, passing through mixed active-sterile
transitions $(0< sin^{2} \xi <1)$. Replacing $N_{n}(x)$ by
$-N_{n}(x)$, we can also get the effective Hamiltonian for
relevant antineutrinos.

Now we consider the uncertainty in Earth matter and its
implications on the atmospheric neutrino oscillations. In some
simple cases, for example, if the neutron number density
$N_{n}(x)$ suffers from a global shift (independent of $x$), the
induced effects on $\sin ^{2} \xi$ and $N_{n}(x)$ are degenerate
in the effective Hamiltonian (\ref{ham}) , such as $\sin ^{2} \xi
\rightarrow \frac{1}{(1 \pm 5 \%)} \sin ^{2} \xi$ when $N_{n}(x)
\rightarrow (1 \pm 5 \%) N_{n}(x)$.

Generally, at a given point $x$ in Earth, the available matter
density, which determines the neutron number density, is an
average value with some prescribed errors, such as the widely used
PREM model\cite{prem}. We can define the average density
$\hat{\rho}(x)$ as an average over all samples of density profiles
$\{\rho(x)\}$
\begin{equation}
\hat{\rho}(x) \equiv \langle \rho(x)\rangle =\int \mathcal{D}[\rho
(x)]F[\rho(x)] \rho(x)\ ,
\end{equation}
and the error $\sigma(x)$ as a variance function
\begin{equation}
\sigma(x) \equiv
\sqrt{\langle\rho^{2}(x)\rangle-\langle\rho(x)\rangle^{2}}\ ,
\end{equation}
where $F[\rho(x)]$ is the probability density of the density
sample $\rho(x)$. Accordingly, the averaged probability for the
$\alpha$ flavor neutrino oscillating into the $\beta$ flavor
neutrino should be
\begin{equation}
\label{op} \langle P_{\nu_{\alpha} \rightarrow
\nu_{\beta}}(L,E_{\nu}) \rangle \equiv \int
\mathcal{D}[\rho(x)]F[\rho(x)]P_{\nu_{\alpha} \rightarrow
\nu_{\beta}}(L,E_{\nu},\rho(x),Y_{e}(x))\ .
\end{equation}
with $L$ the neutrino's travelling distance in Earth. Furthermore,
we can write the variance as
\begin{equation}
\label{dop} \delta P_{\nu_{\alpha} \rightarrow
\nu_{\beta}}(L,E_{\nu}) \equiv \sqrt{ \int \mathcal{D}[\rho
(x)]F[\rho (x)](P_{\nu_{\alpha} \rightarrow
\nu_{\beta}}(L,E_{\nu},\rho (x),Y_{e}(x))- \langle P_{\nu_{\alpha}
\rightarrow \nu_{\beta}}(L,E_{\nu}) \rangle)^{2}}\ .
\end{equation}

In this brief report, we introduce a logarithmic normal
distribution\cite{tarantola} to represent the probability density
function of the Earth matter density samples
\begin{equation}
\label{weight} F[\rho(x)]=\frac{1}{\rho(x)\sqrt{2 \pi s^{2}(x)}}
\exp(-\frac{\ln ^{2}[\rho(x)/ \rho_{0}(x)]}{2s^{2}(x)})\ ,
\end{equation}
\begin{equation}
\label{sx} s(x)=\sqrt{\ln[1+r^{2}(x)]}\ ,
\end{equation}
\begin{equation}
\label{rho0} \rho _{0}(x)=\hat{\rho}(x)
\exp[-\frac{1}{2}s^{2}(x)]\ ,
\end{equation}
where $r(x)=\sigma(x)/ \hat{\rho}(x)$ characterizes the precision
of Earth matter density. And then we use Monte Carlo calculations
to generate the values of $F[\rho(x)]$ between $0$ and $1$ at a
given point $x$ along the propagating path of the neutrinos. With
the chosen $\hat{\rho}(x)$ and $r(x)$, we obtain the value of
$\rho(x)$ from Eq.(\ref{weight}) by computing the values of $s(x)$
in Eqs.(\ref{sx}) and $\rho_{0}(x)$ in Eq.(\ref{rho0}). Hence, the
averaged oscillation probability (\ref{op}) and the corresponding
variance (\ref{dop}) can be calculated. Specifically, we take
$Y_{e}(x)=0.5$ and $r(x)=5 \%$ with $ \hat{\rho}(x)$ given by the
PREM in our numerical calculations.

Usually, the experiment results are reported as the event number,
which can be calculated as
\begin{displaymath}
N=\int ^{E_{u}}_{E_{l}} N_{0} \phi _{\nu_{\alpha}}(E_{\nu})\langle
P_{\nu_{\alpha} \rightarrow \nu_{\beta}}(L,E_{\nu})\rangle \sigma
_{\beta}(E_{\nu})C dE_{\nu}
\end{displaymath}
\begin{equation}
\label{na}
 = \sum _{k=1}^{n}N_{0} \phi
_{\nu_{\alpha}}(E_{\nu}^{k})\langle P_{\nu_{\alpha} \rightarrow
\nu_{\beta}}(L,E_{\nu}^{k})\rangle \sigma _{\beta}(E_{\nu}^{k})C
\Delta E_{\nu}^{k}\ .
\end{equation}
where $E_{u}$, $E_{l}$ denote the upper bound and lower limit of
the detecting energy, $N_{0}$ is a normalization factor with unit
conversions, $ \phi _{\nu_{\alpha}}$ is the $ \alpha$ flavor
neutrino beam flux spectrum, $ \sigma _{\beta}$ is the charged
current cross section of $ \beta$ flavor neutrino, $C$ is the
product of the detector's size and running time, and $\Delta
E_{\nu}^{k}$ is the k$th$ energy bin size. Accordingly, we can
define
\begin{displaymath}
\delta N =\int ^{E_{u}}_{E_{l}} N_{0} \phi
_{\nu_{\alpha}}(E_{\nu}) \delta P_{\nu_{\alpha} \rightarrow
\nu_{\beta}}(L,E_{\nu}) \sigma _{\beta}(E_{\nu})C dE_{\nu}
\end{displaymath}
\begin{equation}
\label{nv}
 = \sum _{k=1}^{n}N_{0} \phi
_{\nu_{\alpha}}(E_{\nu}^{k}) \delta P_{\nu_{\alpha} \rightarrow
\nu_{\beta}}(L,E_{\nu}^{k}) \sigma _{\beta}(E_{\nu}^{k})C \Delta
E_{\nu}^{k}\ ,
\end{equation}
\begin{equation}
\label{dnv} r_{N}=\frac{\delta N }{N}\ ,
\end{equation}
as the variance and relative variance of event number caused by
the uncertainty in Earth matter density, respectively.

For example, in the numerical calculations we take the relevant
data listed in the table \textrm{II}, \textrm{IV}, \textrm{V} of
\cite{honda} and table \textrm{III} of \cite{gaisser}. We use the
mass and mixing parameters: $ \Delta m^{2}=2.0 \times 10^{-3}
\textrm{eV}^{2}$, $\sin ^{2}2\theta =1.0$, $ \sin ^{2} \xi
=0.19$\cite{nakaya}, and neutrino travelling distance $L=l-2r \cos
\Theta $, where $l$ is the neutrino production height in
atmosphere(slant distance in km, listed in\cite{gaisser}),
$r=6371$km is the Earth radius and $ \cos \Theta $ is the zenith
angle. We plot the event number produced by the atmospheric muon
neutrinos and antineutrinos in six zenith angles with three energy
bins as well as their variances arising from the uncertainty of
Earth matter density by using Eq.(\ref{na}) and (\ref{nv}). As
shown in Fig.1, we find the variance of event number is smaller
than $ \sim 10^{-3}$, accordingly the relative variance is smaller
than $ \sim 10^{-4}$, for the longest baseline $ \cos \Theta
=-1.0$ with the largest matter effect, while in the other
baselines, the variance and also the relative variance are much
smaller. Hence we can draw a conclusion that the uncertainty of
Earth matter density seems to have no significant effects in the
oscillating scenario of $ \nu_{\mu} \rightarrow \nu_{\tau} cos \xi
+ \nu_{s} sin \xi $.

\begin{figure}
\includegraphics[scale=1.4]{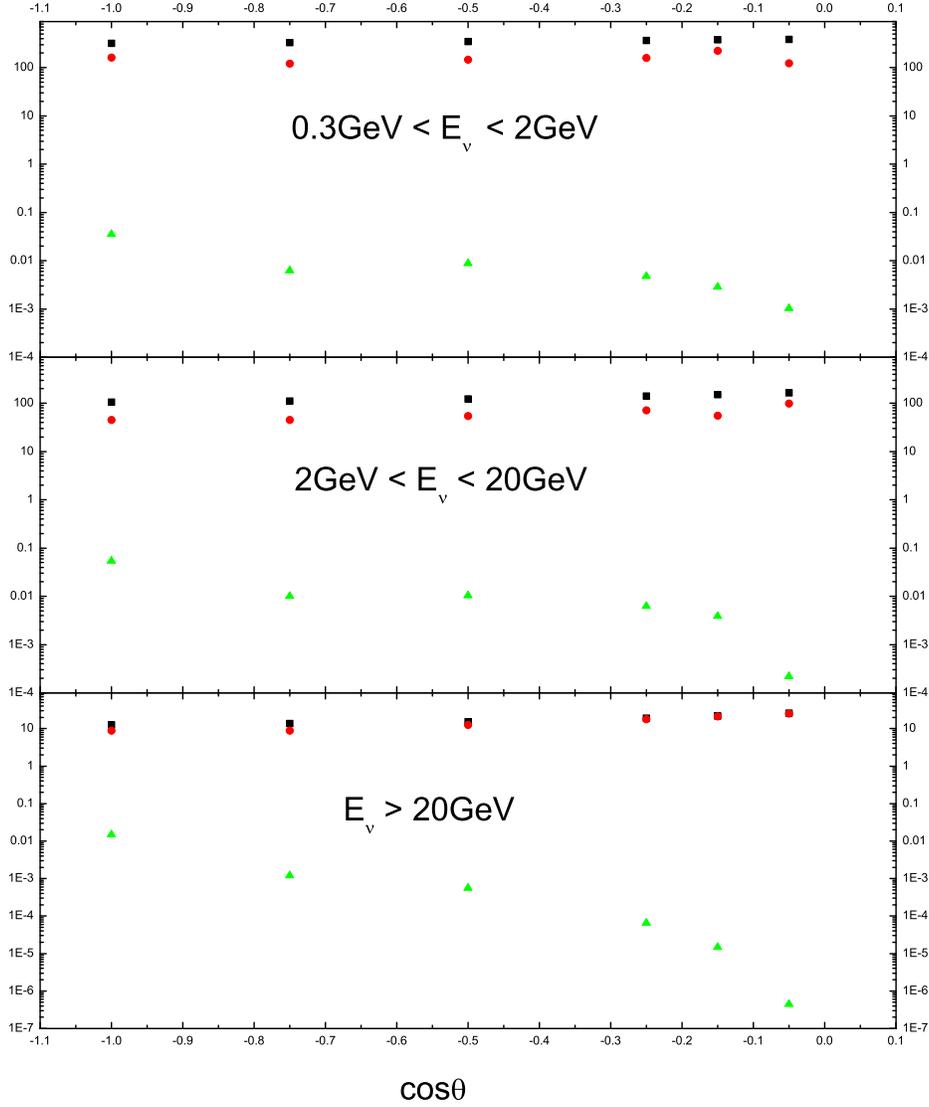}
\caption{Event numbers plotted as a distribution with zenith angle
$ \cos \Theta $ in three neutrino energy bins. The square is
Honda\cite{honda} expectation for no oscillations. The dot is
given by substituting $ \langle P \rangle $ for $P$ in Eq.(8) and
up-triangle represents the variance defined in Eq.(9). These plots
use the mass and mixing parameters: $ \Delta m^{2}=2.0 \times
10^{-3} \textrm{eV}^{2}$, $\sin ^{2}2\theta =1.0$, $ \sin ^{2} \xi
=0.19$\cite{nakaya}, and neutrino travelling distance $L=l-2r \cos
\Theta $, where $l$ is the neutrino production height in
atmosphere(slant distance in km, listed in\cite{gaisser}) and
$r=6371 \textrm{km}$ is the Earth radius.}
\end{figure}

In summary, considering that the Earth matter density has
uncertainty and this uncertainty has significant effects in some
neutrino oscillation cases, such as the $CP$ violation in very
long baseline neutrino oscillations and the day-night asymmetry
for solar neutrinos, we study this uncertainty in the atmospheric
neutrino oscillating scenario of $ \nu_{\mu} \rightarrow
\nu_{\tau} cos \xi + \nu_{s} sin \xi $, and analyze the effects
caused by this uncertainty on the previous fitting results. We
find that this uncertainty seems to have no significant effects
and need not to modify the previous fitting results fortunately.

{ \bf Acknowledgment:} We thank Lian-Lou Shan, Kerry Whisnant,
Bing-Lin Young and Xinmin Zhang for helpful discussions. We also
thank Eligio Lisi for kind comments and suggestions. This work is
supported partly by the National Natural Science Foundation of
China under the Grant No. 90303004.

\end{document}